# The Pollution Effect: Optimizing Keyword Auctions by Favoring Relevant Advertising

Greg Linden
Geeky Ventures[1]
Seattle, WA 98112
glinden@gmail.com

Christopher Meek
Microsoft Research
Redmond, WA 98052
meek@microsoft.com

Max Chickering
Microsoft
Redmond, WA 98052
dmax@microsoft.com

## ABSTRACT
Most search engines sell slots to place advertisements on the search results page through keyword auctions. Advertisers offer bids for how much they are willing to pay when someone enters a search query, sees the search results, and then clicks on one of their ads. Search engines typically order the advertisements for a query by a combination of the bids and expected clickthrough rates for each advertisement. In this paper, we extend a model of Yahoo's and Google's advertising auctions to include an effect where repeatedly showing less relevant ads has a persistent impact on all advertising on the search engine, an impact we designate as the *pollution effect*. In Monte-Carlo simulations using distributions fitted to Yahoo data, we show that a modest pollution effect is sufficient to dramatically change the advertising rank order that yields the optimal advertising revenue for a search engine. In addition, if a pollution effect exists, it is possible to maximize revenue while also increasing advertiser, and publisher utility. Our results suggest that search engines could benefit from making relevant advertisements less expensive and irrelevant advertisements more costly for advertisers than is the current practice.

## Categories and Subject Descriptors
J.4 [**Computer Applications**]: Social and Behavioral Sciences – *Economics*

## 1. INTRODUCTION
Microsoft Live Search, Yahoo, Google, and other web search engines sell the right to show advertisements in slots placed on the search results page. Search engines solicit bids from advertisers on how much they are willing to pay for a click on their ad when a searcher is viewing the search results for a list of queries. When a query comes into a search engine, in addition to finding search results for the query, the search engine picks a selection of advertisements to show for the query. Search engines rank order the candidate advertisements by a combination of the amount the advertiser bid and the estimated probability that searchers will click the advertisement, then pick the top candidates as the ads to show.

For example, several advertisers may be interested in clicks by searchers who search for [credit card]. By offering bids on the term [credit card], the advertiser can attract traffic to their website. On a search for [credit card], the search engine will look at all the advertisers who bid on this query, rank order the advertisers primarily by a combination of the bid and how likely searchers appear to be to clicking on the ad for this advertiser, and display a selection of the top ranked ads.

Although advertisers are bidding on clicks, the search engines are selling impressions, the space on the page. Clicks are related to impressions by the clickthrough rate (CTR), which is the probability of clicking on an advertisement when it is displayed. The CTR of an advertisement depends not only on the relevance of the advertisement, but also on its position. All else being equal, ads in higher positions on the page have a higher likelihood of being seen by searchers and attract more clicks.

Historically, Yahoo and Google used different rankings of candidate advertisements. Yahoo ranked candidates in decreasing order of their bids. Google ranked candidates in decreasing order of the bid multiplied by the estimated CTR of the ad (i.e. $Bid \times CTR$). The first approach ranks the ads by the value per click; the second ranks by the value per impression. Currently, Yahoo, Google, Microsoft Live Search, and most other web search engines use value per impression as the primary factor to rank order the candidate advertisements.

Previous work by Lahaie [1] modeled and analyzed a family of ranking rules of the form $Bid \times CTR^\alpha$. This family of rules includes the old Yahoo value per click, the current value per impression model used by most search engines, as well as combinations of the two. In Monte-Carlo simulations that used distributions based on Yahoo bid and clickthrough data for a specific high-volume keyword, this earlier work found that the optimal setting of alpha to maximize revenue was 0, which is equivalent to the older Yahoo model of ordering advertisements by the value of the bids.

In this paper, we reproduce the simulations and results of [1] and then extend the model they used to include what we call the *pollution effect*, the impact of repeatedly showing searchers irrelevant advertisements that do not attract clicks. We argue that the primary reason that we see the pollution effect exists is that searchers who see irrelevant advertisements look at all advertisements less in the future, a phenomenon known as *ad fatigue*.

Our new model that includes a pollution effect is a simple and reasonable extension of the model used in [1]. Our major change from previous work is to make small changes in the CTR distribution as alpha changes. The authors of the earlier work used a CTR distribution derived from empirical data at a specific alpha setting. However, different settings of alpha almost certainly will change the advertisements shown, impact both advertiser and user behavior, and change the CTR distributions. Our model of the pollution effect uses slightly different CTR distributions as alpha is changed.

---
[1] This work was performed while the author was at Microsoft.

Monte-Carlo simulations based on our new model suggest that, for the advertising ranking functions commonly used by search engines, $Bid \times CTR^\alpha$, the optimal setting of alpha is at least 1. This result disagrees with the conclusion of previous work, but confirms the current settings used by major search engines, all of which use an alpha setting of 1. In addition, our results suggest that even higher settings of alpha may increase publisher, advertiser, and user utility simultaneously.

The contributions of our paper are a simple and plausible model for the impact less relevant ads have on other ads over time, simulations using that model that show the ranking function that will optimize revenue, and results that suggest that higher settings of alpha may be optimal. More generally, our work indicates that search engines may be missing a substantial opportunity to increase value for searchers, advertisers, and themselves by not having much higher penalties on advertisements that searchers do not find relevant.

## 2. RELATED WORK

Our work can be seen as building on the work of [1], using a similar model, comparable Monte-Carlo simulations, and their probability distributions derived from Yahoo data. However, our conclusions are dramatically different, suggesting that search engines should heavily favor advertising that searchers find more relevant. We show the conclusions of past work are sensitive to an implicit assumption that overall CTR is independent of the setting of alpha for the ranking function. We relax this assumption using a plausible relationship between overall CTR and the setting of alpha. We show that the new model has different conclusions about the optimal setting of alpha. Our models support the current search engines rank ordering of advertising by $Bid \times CTR^\alpha$ where alpha = 1, but also suggest that alpha > 1 to further favor relevant advertising may lead to higher long-term revenue. Therefore, our contributions to earlier work are a new model that includes the pollution effect and new conclusions about the optimal rank order for advertisements given the existence of the pollution effect.

Lahaie [5] describes a theoretical framework on which his later work [1] was based. Our work also benefits from the framework described in this paper.

The papers of Varian [2] [3], Athey et al. [12], and Edelman et al. [4] lay the foundation for our work. These papers discuss keyword auctions, the models used by Yahoo and Google, how the auctions encourage advertisers to be truthful about their actual value of a click, and how they maximize revenue. One of these papers, Varian [2], includes a section on "ad quality" that states that "showing a 'bad ad' can affect users' future propensity to click." Our work extends Varian's statement with a simple and plausible model of how irrelevant advertisements can affect user's future propensity to click and shows the optimal ranking function for advertisements under that model. In addition to a detailed discussion of auction models, Athey et al. [12] mention that welfare can be hurt by irrelevant advertisements and briefly discuss a detailed model of consumer behavior that includes consumer belief in the quality of a publisher and relevance of advertisements. However, they do not explore the possibility of a drop in welfare in depth.

A few recent papers have touched on the idea that irrelevant advertisements may have a negative impact on other advertisements. Perhaps most related to our work is Ghosh et al. [6] with their discussion of an "externality effect" in advertising where the utility of a winner also depends on the other winners. However, they only look at the impact of an ad on other ads in the same auction, not across time, and are not modeling the idea that an irrelevant ad can have a long-term impact on all other advertisements by causing searchers to look at advertisements less frequently. Broder et al. [13] argue that irrelevant advertisements annoy users and provide no economic benefit, then examine a machine learning approach to learning when not to show advertisements. However, the authors do not model a pollution effect where irrelevant advertisements reduce future clicks on all other advertisements and they leave considering user ad fatigue to future work. Gunawardana et al. [7] consider the negative impact ad aggregators may have on other advertisers, but do not develop this into a more general model. Abrams [14] offer a technique for charging the cost of irrelevant advertisements to search engines back to the advertisers, but ignore the impact of irrelevant advertisements on searchers and other advertisers.

## 3. POLLUTION EFFECT

Our model includes a pollution effect, an externality where advertisements can impact other advertisements in the auction. In particular, we include an effect where each time a searcher sees irrelevant advertisements, they have a greater tendency to assume all future advertisements will be irrelevant, and so have a lower likelihood of viewing them. That is, advertisements that are not relevant cause searchers to have a lower probability of looking at the advertising section of the search result page.

We assume that the pollution effect is uniform across all ads; a searcher either decides to look at the advertising section of the page or ignores the entire section. We also assume the choice to look at the advertising section is independent of the choice to not look at the next ad when reading all the ads, so that it is reasonable to consider the model of position bias independently of the model of the pollution effect.

We choose a simple and abstract model of the pollution effect that merely shifts the CTR distribution for all of the advertisements. This represents searchers looking at the advertisements less frequently, reducing the clicks all advertisements receive uniformly.

In our work, we shift the CTR distribution as alpha changes, taking advantage of the inverse relationship between alpha and the relevance of advertisements shown. To see this relationship, note that since we are ordering the advertisements by $Bid \times CTR^\alpha$, for the same bids, lower values of alpha will cause advertisements with lower CTR to be featured more prominently. Likewise, higher alpha values will tend to move higher CTR ads to better positions. When higher CTR ads have a greater likelihood of appearing in the top slots, they have a higher likelihood of being seen, so the average relevance of the advertisements will be higher. We confirm this relationship between alpha and CTR in our simulations as described later in this paper.

Rather than shift the CTR distribution, another choice may be to attempt to explicitly model the pollution caused by each individual advertisement each time it is shown (e.g. [12]). However, this choice would introduce much complexity and many new parameters when we are primarily interested in aggregate effects in this work. Our model does not capture short-term effects when alpha changes, but a new, shifted CTR distribution will capture the long-term steady state after a change in alpha. Once searchers have adjusted to a given setting of alpha, the CTR distribution describes how people view and click on ads. Our approach to understanding the impact of a pollution effect was to seek the simplest model we could find that could still yield insight on the pollution effect in advertising auction. Shifting the CTR distributions as alpha changes appears to be a simple, plausible, and effective model of the pollution effect.

## 4. POLLUTION TAX

A pollution tax may be an instructive framework for thinking about how to deal with the pollution effect in advertising auctions. A pollution tax attempts to fix an externality by requiring an extra payment to compensate those impacted by pollution that otherwise would not be compensated. In advertising auctions, we could add a penalty that resembles a pollution tax that is intended to make advertisers pay the full cost of showing a bad advertisement. If showing an irrelevant advertisement causes searchers to now look at all ads slightly less frequently, the other advertisers should be compensated for their loss.

Ideally, a pollution tax should match the full cost of the pollution to all others. In an advertising auction, a pollution tax should charge irrelevant advertisements for the cost to all other advertisements, which should be calculated in part based on the drop in likelihood that the searchers will look at the advertising section.

In our model, we use the implied pollution tax from higher settings of the alpha parameter when ordering the advertisements by $Bid \times CTR^{\alpha}$. We assume that when a searcher clicks on an ad, that person found the ad useful, so we effectively are using CTR as a proxy for user utility (in practice, this assumption can be violated, which we discuss further in Section 5.2). Therefore, when ranking by $Bid \times CTR^{\alpha}$, higher values of alpha will charge irrelevant advertisements more. Using Monte-Carlo simulation, we determine the ideal surcharge embedded in the ideal setting of alpha.

Rather than representing a pollution tax in alpha, another choice may have been to explicitly model the pollution tax as a separate term (e.g. [14]), either in aggregate or individually for each advertisement. However, this choice would introduce much new complexity to the model. We sought to use the simplest model possible that could still represent a pollution tax. We also sought to make as few changes as possible to the family of ranking rules $Bid \times CTR^{\alpha}$ that are currently used by most search engines for advertising auctions. Embedding the pollution tax in the alpha parameter introduces no additional complexity while still allowing us to examine a range of advertising ranking functions that make advertisements that searchers find irrelevant more costly.

## 5. MODEL

We examine a family of ranking functions of the form $Bid \times CTR^{\alpha}$. When alpha is 0, the advertisements are ordered by bid, which is equivalent to the old Yahoo model. When alpha is 1, the advertisements are ordered by value per impression, which is the ranking function currently used by most search engines. Alpha values in the range [0,1] yield models that are combinations of the old and new Yahoo ranking functions, and so weight bids more heavily than is common.

Alpha values above 1 penalize advertisements that have low clickthrough rates (or, equivalently, subsidize advertisements that searchers click on frequently) and are not common in practice. Since we reproduce the results of previous work [1] as well as extending it, we explore negative alpha values in the range [-2, 0] as the earlier work did, but we should note that those ranking functions are usually considered impractical given that they have an undesirable effect of strongly favoring advertisements that get few clicks.

We model the advertising placements as K slots for advertisements for which there are N bidders. We assume that K < N for the purposes of the model and the simulations. This assumption does not limit the generality of our results since auctions with fewer advertisers than slots could fit the model as a reserve price beyond which the placement is more valuable to the search engine left empty or as advertisers paying nothing for the placement.

Position bias is the tendency of searchers to view and click advertisements in lower positions less often than advertisements in higher positions. We assume the position bias on the advertisements – the tendencies of ads in lower slots to be seen and clicked less frequency – is independent of the advertisements and the query. Therefore, we assume the CTR of an ad can be estimated independently of its position and assume that unbiased CTR is available for our ranking function. We label the position bias at slot $i$ in the advertisements as $x_i$, label the CTR of the ad in slot $i$ as $CTR_i$, and define the actual CTR an ad gets in slot $i$ as $CTR_i \times x_i$.

We model generalized second price auctions, so we set the advertiser payments for a click not at the bid, but at the lowest bid necessary to maintain the position. For example, if there are two advertisements, one with a bid of $3, the other $2, and both have a CTR of 1%, then the first advertisement would get the top position, but only pay $2, not $3. In general, the amount the advertiser at slot $i$ pays for an ad is $Bid_{i+1} \times \frac{w_{i+1}}{w_i}$, where $i+1$ is the advertisement that wins the slot immediately below advertisement at slot $i$ and $w_i = (CTR_i)^{\alpha}$.

We assume that bidders are rational and are playing as best they can given the other bids and bidders in the auction. As in [1] and as supported by [3], we assume that bidders are playing the smallest symmetric equilibrium, which causes bidders to be ranked in decreasing order of $w_i \times Value_i$, where $w_i = (CTR_i)^{\alpha}$ and $Value_i$ is the unknown true value of a click to the bidder. For further discussion of symmetric equilibrium (which is sometimes referred to as "locally envy-free equilibrium") and truth telling in advertising auctions, we refer the reader to [1], [2], [3], [4], and [12].

We assume exact match for the keyword auctions, meaning that advertisers must bid on the query exactly to be eligible for a slot. In practice, search engines do fuzzy matching between the keywords bid and the search query terms, sometimes even showing the advertisements for different queries than the bidded terms if they appear to be semantically or behaviorally related.

However, this added complication would add nothing to our model nor would it change our conclusions.

## 5.1 Revenue

We assume the goal of the search engine is to maximize long-term revenue from its advertisements. Therefore, the search engine should choose a ranking function that accomplishes this goal.

Since these are generalized second price auctions, the price actually paid for an ad is the minimum bid necessary to maintain position, or $Bid_{i+1} \times \frac{w_{i+1}}{w_i}$, where $w_i = (CTR_i)^\alpha$, as described earlier. This price is only paid when the ad is clicked, so the revenue for a single ad will be $(CTR_i \times x_i) \times Bid_{i+1} \times \frac{w_{i+1}}{w_i}$. The revenue for all the ads on a search result page, therefore, will be

$$\sum_{i=1}^{K} (CTR_i x_i) Bid_{i+1} \frac{w_{i+1}}{w_i}$$

And, the total revenue for the search engine overall will the sum of the revenue of each search result page.

Although advertisers do not necessarily bid their true value, we can state the revenue as a function of advertiser value when the bids stabilize under the assumption of symmetric equilibrium. The revenue for one ad on a search result page will be

$$CTR_i \times x_i \times Bid_{i+1} \times \frac{w_{i+1}}{w_i} = \sum_{j=i}^{K} CTR_i (x_j - x_{j+1}) Value_{j+1} \frac{w_{j+1}}{w_i}$$

where K is the number of ads in the auction. Intuitively, the bid of an advertiser depends on the actions of all the other advertisers, but the bids in the auction will settle into a steady state that depends on the value each the advertisers place on a click.

The total revenue of an auction, stated as a function of advertiser value, will then be

$$\sum_{i=1}^{K} \sum_{j=i}^{K} CTR_i (x_j - x_{j+1}) Value_{j+1} \frac{w_{j+1}}{w_i}$$

And, the total revenue for the search engine overall will be the sum across all auctions.

Please see [1] for further discussion of symmetric equilibrium and the derivation of revenue as a function of advertiser value.

## 5.2 Relevance

Relevance is a measure of the utility of the advertisements to searchers. Advertisements often contain information about products and services that may be useful to the person executing the search.

A search engine needs to balance the needs of searchers against its own goals. If searchers constantly find the advertisements irrelevant and uninformative, they will stop looking at them [9].

In this work, we use CTR as a proxy for user utility. The assumption is that when a searcher clicks on an ad, that person found the ad useful. In practice, this assumption can be violated by deceptive advertising; we assume that techniques to detect and eliminate clicks on deceptive ads – such as dropping clicks where the searcher leaves the advertising site immediately – are already part of our CTR. We also assume that we can estimate CTR accurately even for new ads. In practice, accurately estimating clickthrough rate given limited actual click data is a hard problem (see, for example, [10]).

Therefore, we define the relevance of an ad in slot $i$ as its actual clickthrough rate, $CTR_i \times x_i$, where $x_i$ is the position bias for slot $i$. The total relevance of the advertisements on a search result page will be

$$\sum_{i=1}^{K} CTR_i x_i$$

and will vary depending on the ranking function because which ads are selected for which positions change with the ranking function. The total relevance for the search engine is the sum of the relevance of each search result page.

Previous work (e.g. [1]) assumed that the relevance distribution of the ads is independent of the setting of alpha. In our model, we change the relevance distribution for the ads as alpha changes. In this way, our model includes an effect where, when the advertising section of a page becomes less relevant to searchers, searchers start looking at all the ads less frequently.

## 5.3 Efficiency

Efficiency is a measure of advertiser satisfaction. It is the total value an advertiser receives from the auction. Efficiency for a single advertiser is $CTR_i \times x_i \times Value_i$. Efficiency of the advertisements in all slots on a search result page will be

$$\sum_{i=1}^{K} CTR_i x_i\, Value_i$$

The total efficiency of the search engine will be the sum of the efficiency of each search result page.

The efficiency can be interpreted as the revenue paid to the search engine plus the excess value the advertiser keeps in a generalized second price auction from the minimum necessary to maintain their position rather than their bid. This is easy to see if we break the efficiency the first equation above into two terms, one for the revenue for the search engine and one for the excess value retained by the advertiser

$$CTR_i \times x_i \times Value_i = CTR_i \times x_i \times p_i + CTR_i \times x_i \times (Value_i - p_i)$$

where $p_i$ is the price paid for the for the advertisement, so $p_i = Bid_{i+1} \times \frac{w_{i+1}}{w_i}$.

## 6. SIMULATIONS

To determine the impact of different settings of alpha, we created a Monte-Carlo simulation of millions of advertising auctions, then examined the impact of the setting of alpha on revenue, relevance, and efficiency. We then created a dependency between alpha and the relevance distribution for the advertisements to determine how sensitive the optimal setting of alpha is to an assumption of independence.

In the Monte-Carlo simulation, for additional realism, we used the marginal distributions from [1] that the authors derived from empirical Yahoo bid and clickthrough data for a high volume keyword. Specifically, as in the previous work, we used a lognormal distribution fitted to the empirical data for the value of a click (mean of 0.35 and stddev of 0.71), and a beta distribution fitted to the CTR data (alpha of 2.71, beta of 25.43).

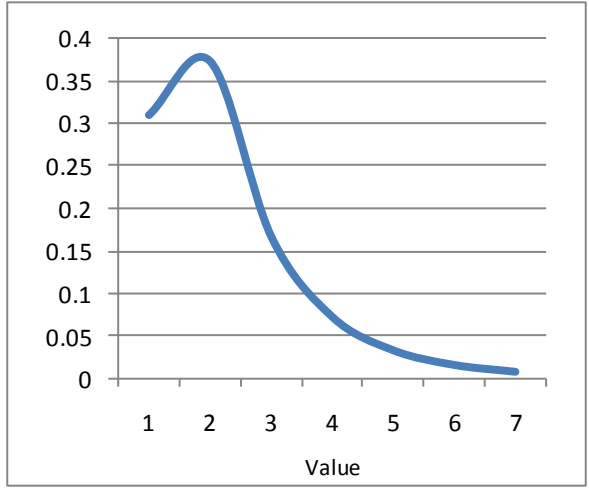

**Figure 1: Lognormal marginal distribution of value derived from empirical Yahoo data.**

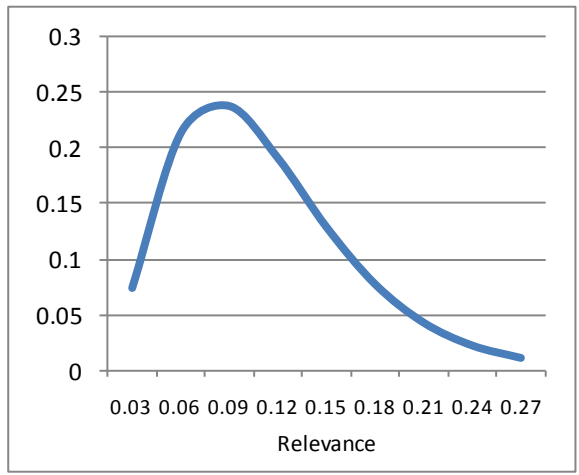

**Figure 2: The beta marginal probability distribution for the relevance of an advertisement derived from empirical Yahoo data.**

Our Monte-Carlo simulation was for a total of 9.6 million auctions of search result pages with 12 slots and 13 bidders. We drew advertiser bids the CTR of the ad randomly for each of the 13 advertisers from the distributions above, then simulated an auction for the advertisers, ranking them by $Bid \times CTR^\alpha$, then computing the revenue, relevance, and efficiency. We repeated this for all alpha in the range [-2,2] in 0.1 increments, calculating the total revenue, relevance, and efficiency for each setting of alpha.

The empirical data from [1] showed a positive Spearman correlation of 0.4 between the value of an advertisement to an advertiser and the CTR of the ad. We used a Gaussian copula – a method from finance of creating a joint distribution from two marginal distributions – to allow the Monte-Carlo simulation to represent this dependence. We simulated a wide range of possible correlations between value and relevance, but focused our attention on the most realistic correlation of 0.4.

We modeled position bias using the positional effects observed in Yahoo empirical data from [1]. We choose using the Yahoo data over a simpler model of bias both to be able to replicate the results of previous work and to increase the realism of our simulations.

Before we attempted to analyze the impact of the pollution effect, we ran our simulations using these parameters, which we sought to make identical to the parameters used in [1]. Our goal was to reproduce the results in previous work before extending their model. Our results, presented in the next section, show that these simulations yielded the same results and conclusions as the earlier work. In particular, our results also showed a setting of alpha of 0 was optimal.

Finally, we consider an aggregate model for the pollution effect. Previous work assumed independence between alpha and the relevance distribution. Our concern is that the Yahoo empirical data was taken from advertising auctions run with one particular setting of alpha. We suspected that the empirical data and hence the relevance distribution would be quite different if taken from advertising auctions with a different setting of alpha.

For intuition on why this might be the case, take the example of where alpha is set to -1. This setting of alpha charges less to advertisements that do not get clicks. The advertisements winning the top slots likely would be advertisements on which searchers do not click. After being shown these advertisements repeatedly, searchers likely would start perceiving all the advertisements as irrelevant and stop looking the ads.

We represented the pollution effect by slightly shifting the relevance curve as alpha changes. Specifically, we set the beta parameter of the relevance beta distribution at 25.43, as derived in [1] for empirical data when alpha was 1, but then we change the beta parameter over the range [18.43,46.43] as alpha changes in the range [-2, 2]. This smoothly shifts the CTR distribution to the left as alpha drops to 0 and below, representing the decreasing likelihood that searchers will view and click on ads after continually being shown irrelevant ads. Likewise, the CTR distribution shifts slightly to the right as alpha increases beyond 1, representing the increasing likelihood that searchers will view and click on ads after repeatedly being shown useful ads.

With this simple model, we are able to simulate what happens if the auction makes it easier for less relevant ads to appear. Searchers are likely to suffer ad fatigue and look advertisements less frequently when they see irrelevant ads. If the search engine makes it less expensive for irrelevant ads to appear and starts showing irrelevant ads more prominently, it will cause ad fatigue. We call this the pollution effect of irrelevant ads. Our model represents the pollution effect by shifting the relevance distribution.

The chart below shows the shifts in the relevance distributions that we used in our simulations for alpha in the range [0, 1.5].

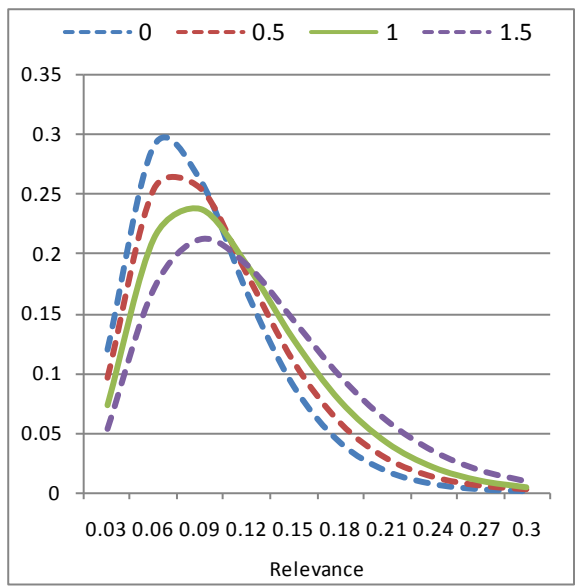

**Figure 3: Relevance distribution curves as alpha changes from [0,1.5]. The green line for alpha = 1 is the original relevance distribution.**

The shifts in the CTR distribution are quite modest over the realistic range of alpha of [0, 1.5]. With this simple model, we can capture differences in how searchers view and click on advertisements when they are shown more or fewer irrelevant advertisements. We believe that this shift in the CTR distribution is a reasonable and conservative approximation of the pollution effect that search engines see in practice.

## 7. RESULTS

We ran two groups of simulations, the first with an assumption that the relevance distribution for the ads remains constant regardless of the ads users see, the second designed to relax the assumption of independence between alpha and the relevance distribution and examine the impact of one plausible shift in the relevance distribution as alpha changes. The first was intended to reproduce the results of previous work, the second to extend previous work to include the pollution effect.

The results from the first group of simulations establishes that the optimal setting of alpha to maximize revenue is 0 when ranking advertisements by $Bid \times CTR^{\alpha}$. This confirms the results of previous work [1] where their simulations also found an optimal setting of alpha at 0. Figure 4 below shows total revenue normalized to the range [0,1] for alpha in the range [-2,2] using Yahoo's empirically observed correlation of 0.4 between the value distribution and the relevance distribution.

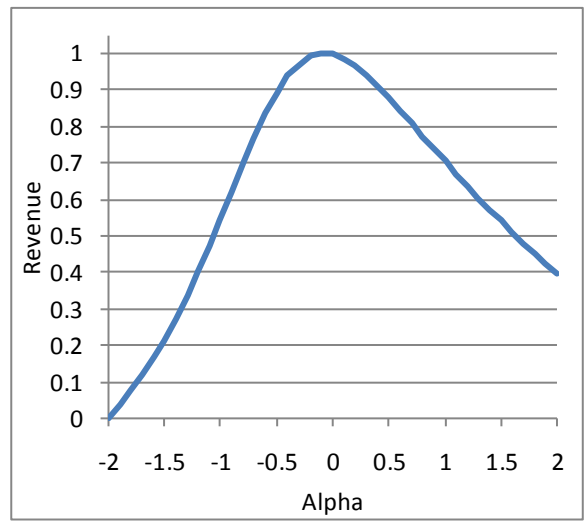

**Figure 4: Normalized total revenue for alpha in [-2,2].**

The revenue peaks when alpha is at 0. This means that revenue is maximized when we order advertisements almost entirely by their bids. Note that this is equivalent to the old Yahoo auctions where they would order by value per click, but that all major search engines currently order their advertisements by value per impression (alpha = 1).

Total efficiency for the auctions is shown below.

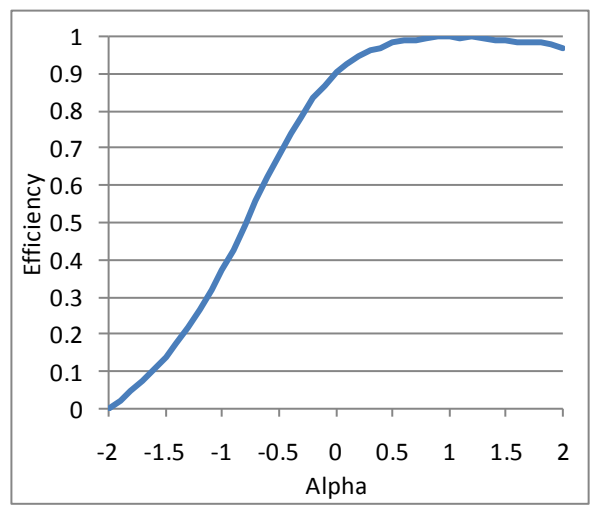

**Figure 5: Normalized total efficiency for alpha in [-2,2].**

Recall that total efficiency is the sum of the value to the search engine and to the advertiser. It is interesting that efficiency is essentially flat over the range [0, 2], meaning that the total value created and shared by advertiser and search engine is the same over a wide range.

Total relevance for the auctions is shown below. While total revenue (in Figure 4) peaked at alpha of 0, the total relevance shows a much different pattern.

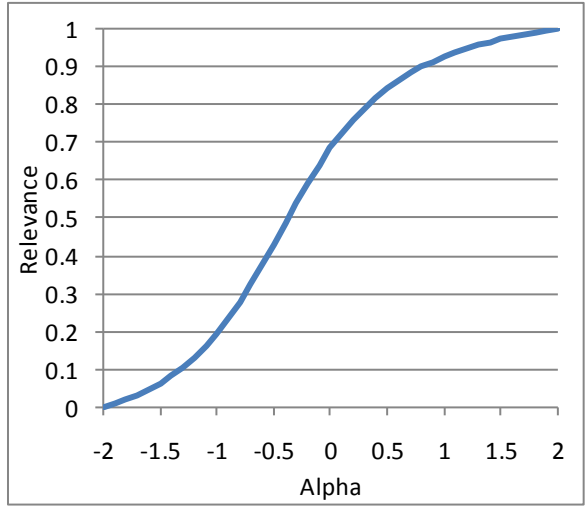

**Figure 6: Normalized total relevance for alpha in [-2,2].**

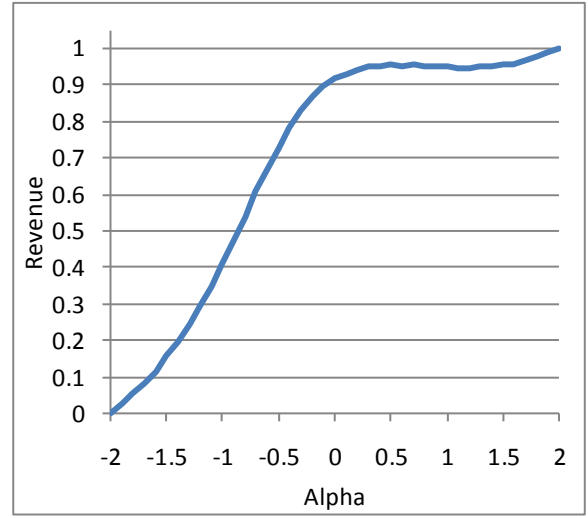

**Figure 7: Normalized total revenue for alpha in [-2,2], with pollution effect, CTR distribution changing with alpha.**

This graph makes the relationship between alpha and relevance clear. Relevance increases as alpha increases. This result confirms our earlier claim that a strong relationship exists between alpha and relevance. It suggests that the irrelevant ads will be shown more prominently as alpha decreases and the overall CTR rates for the auctions will change. In our second round of simulations, we will use a model that shifts the relevance distribution as alpha changes.

Looking at all three graphs, as in the previous results, it appears we can optimize revenue while also nearly optimizing efficiency with a setting of alpha near 0. However, the same cannot be said of relevance. It seems we cannot make the search engine happy without hurting searchers; we cannot maximize the usefulness of the advertisements to users without impacting revenue.

Our results from our second round of simulations show that this tradeoff disappears once we include the pollution effect in our model. The new model shifts the CTR distribution as alpha changes. The old model used in previous work assumed that the empirically observed relevance distribution from an alpha of 1 would apply to other settings of alpha. As we have argued and as results from the simulations show, changing alpha will make it easier for irrelevant ads to appear and likely will cause searchers to look at all advertisements less frequently. Our new model includes this effect.

In the second round of simulations, we slightly changed the mean of the CTR distribution for the advertisements as described earlier. Our results from this group of simulations show that alpha in the range [0.3, 2.0] now optimizes revenue.

The revenue now appears to be essentially flat a broad range of alpha settings. A detailed look at the data shows the maximum to be at alpha 1.5 and above, but that any value in the range [0.3, 2.0] is close to maximizing revenue.

With the new model, no longer do we see a sharp peak near alpha of 0. In fact, it appears that the search engine could be essentially indifferent to a wide range of alpha values, perhaps allowing alpha to be determined based on other factors.

The following graph compares the revenue as alpha changes under the new model with the pollution effect and old model from previous work.

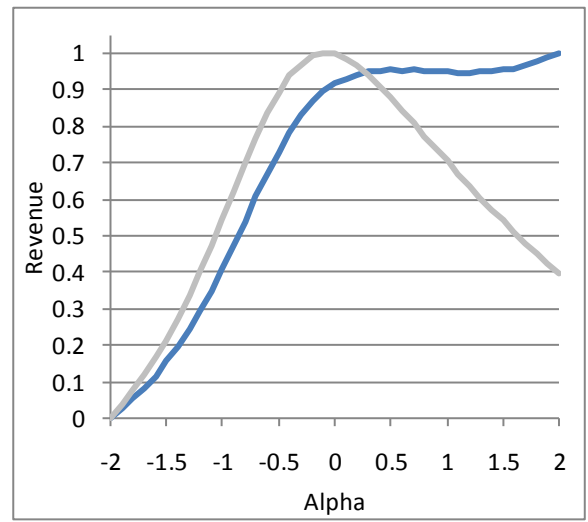

**Figure 8: Normalized total revenue for alpha in [-2,2] comparing revenue curves with the pollution effect (in blue) and without (in gray).**

Total efficiency and total relevance for the second round of auctions are shown below. Note that the total relevance and total efficiency keep increasing as alpha increases.

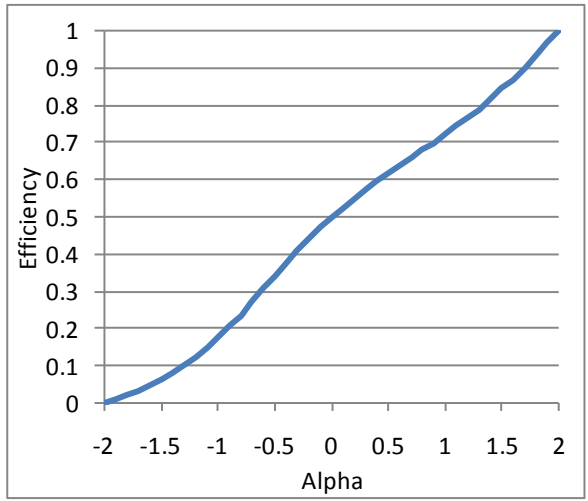

**Figure 9: Normalized total efficiency for alpha in [-2,2] with pollution effect, CTR distribution changing with alpha.**

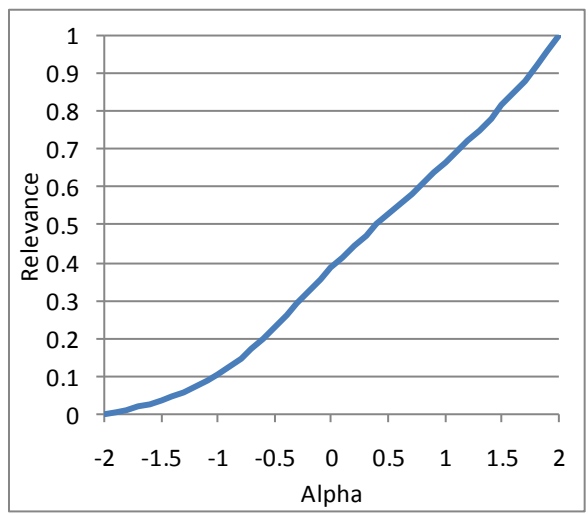

**Figure 10: Normalized total relevance for alpha in [-2,2] with pollution effect, CTR distribution changing with alpha.**

Once we included a pollution effect, both efficiency and relevance increase nearly linearly as alpha increases. The higher the alpha, the more combined value is created for advertisers and the search engine (efficiency) and the more searchers like the advertisements (relevance).

Higher settings of alpha mean that advertisements with higher CTR will get higher position for the same bid. In essence, relevant advertisements are being given a discount to maintain the same position. If relevant advertisements maintain their bids, they will tend to be featured more prominently and seen more frequently. Likewise, irrelevant advertisements require higher bids to maintain their position and therefore will tend to be shown less frequently. Because the average advertisement will be more relevant to searchers, searchers will tend to look at the advertisements more frequently and click on them more often.

As a result, overall value the search engine creates for advertisers and the search engine combined increases. However, the search engine revenue does not increase since the search engine is discounting relevant advertising to attract more searcher attention to the advertising.

Considering the impact to revenue alone, the search engine should be mostly indifferent to a wide range of alpha between [0.3, 2.0]. But, given the search engine's indifference, higher alpha settings in the range may be attractive since they create additional value for searchers and advertisers.

Finally, we should note that we ran additional simulations to test the sensitivity of our model to some of the parameters. In particular, we looked at a wide range of variations in the correlation between value and relevance as well as several smaller variations in how much the pollution effect shifts the CTR distribution of the advertisements as alpha changes. We discovered that the results are not sensitive to small variations in the correlation between value and relevance, but are somewhat sensitive to variations in the pollution effect.

For example, if the shifts in the CTR distributions as alpha changes are 20% lower than what we assumed, the optimal settings of alpha to maximize revenue drops to the range [0.2, 0.5], peaking at 0.4, but allowing settings as high as 1.1 with only a modest loss in revenue. In the other direction, we found that if the shift in the CTR distributions are 20% higher, then the revenue curve in Figure 7 changes to slope upward beyond an alpha of 1, suggesting that alpha in the range [1,2] is optimal. These shifts in the revenue curves from varying the magnitude of the pollution effect are shown in Figure 11 below.

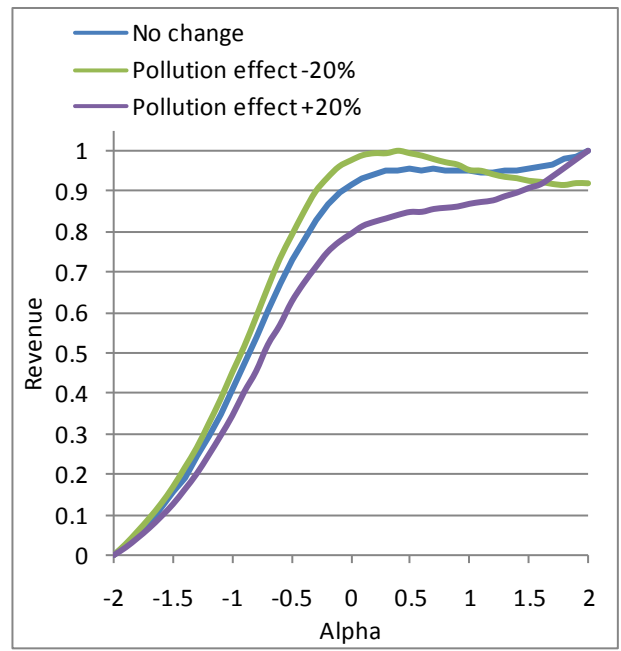

**Figure 11: Normalized total revenue for alpha in [-2,2] showing the sensitivity of the total revenue to magnitude of the shift in the CTR distribution as alpha changes.**

The modest sensitivity of the model to the magnitude of the pollution effect suggests a need for more empirical data on the exact nature of the effect, especially at higher alpha settings. However, one can still conclude both that a pollution effect has a substantial impact on the choice of ranking function and that reasonable models of the pollution effect suggest using alpha of at least 1 for the ranking function.

## 8. DISCUSSION

The goal of our work is to examine the impact of the pollution effect. We suspect that bad advertisements not only impact themselves, but all other advertisements, primarily by causing people who see irrelevant advertisements to reduce their likelihood of looking at advertisements in the future.

We wanted to use as simple of a model as possible that still captured information we have from empirical data about typical advertising auctions and how searchers behave in advertising auctions. We sought a model of the pollution effect that would not require changes to how search engines currently rank advertisements in their auctions, but would still show the impact irrelevant advertisements have on searcher behavior and other advertisers. The model we used represents the pollution effect with modest shifts in the CTR distribution as the alpha setting of ranking functions of the form $Bid \times CTR^\alpha$.

Our most important results for our new model are represented in Figures 7-10. These results are for the most realistic settings. We used the relevance distribution, value distribution, and correlation between value and relevance distribution that was observed in Yahoo data. We then shifted the relevance distribution as alpha changed to model the pollution effect. The results show that revenue is optimized with any alpha in the range [0.3, 2.0], but that higher alpha values in that range yield higher relevance and efficiency, benefiting searchers and advertisers.

Given that most search engines currently order by $Bid \times CTR$ – using an alpha of 1 – this result is surprising. It suggests that most search engines may be missing an opportunity to improve the usefulness of their advertising to searchers and advertisers without any long-term impact to revenue.

To further understand the importance of modeling the pollution effect, it is also instructive to compare the results in Figures 7-10 with Figures 4-6. Figures 4-6 reproduce the previous work and show the total revenue, efficiency, and relevance without modeling the pollution effect; Figures 7-10 include the pollution effect. The conclusion in previous work about the optimal setting of alpha is quite different. If we assume that relevance remains unchanged as alpha changes, as previous work did, the results are that an alpha near 0 optimizes revenue. However, once we recognize that lower settings of alpha will tend to show irrelevant ads and will make searchers view and click on advertisements less frequently – the pollution effect – we find much higher values of alpha optimize revenue. In addition, we find that it is now possible to optimize revenue, efficiency, and relevance simultaneously with setting of alpha at or above 1.

Although at least one major search engine used an alpha of 0 in the past, our new model confirms that the higher alpha setting of 1 now commonly used is much closer to optimizing revenue. It is possible that search engines discovered experimentally that a higher alpha setting does optimize long-term revenue; this would explain why almost all search engines now use the higher setting. We take this agreement between the conclusions of our model and what is done in practice as evidence for the importance of modeling the pollution effect and as support for the validity of our model of the pollution effect.

The conclusions of our model go beyond what search engines currently implement, suggesting that an alpha settings above 1 may be desirable. Some search engines recently appear to be experimenting with higher penalties for irrelevant advertising – Google's Ad Quality [11] measure could be seen as a increasing the importance of relevance when ranking ads and thereby charging irrelevant advertisements more – and, if the conclusions of our model are a guide, we expect their experiments to be fruitful.

## 9. CONCLUSION AND FUTURE WORK

Our model of the pollution effect suggests that it is possible to simultaneously maximize value for search engines, searchers, and advertisers by favoring relevant advertisements and penalizing less relevant advertisements.

The primary conclusion of our work is that the pollution effect can dramatically change which ranking rule yields optimal revenue for a search engine. When ranking advertisements by $Bid \times CTR^\alpha$, an alpha setting of 0 appears to be optimal without a pollution effect, but a wide range of alpha settings between [0.3, 2.0] equally maximize revenue with the pollution effect. Moreover, with a pollution effect, settings of alpha at 1 or higher also appear to maximize the relevance of the advertising for searchers and the efficiency of the advertising for advertisers.

Most search engines currently use an alpha of 1 when ranking advertisements. Our work suggests that higher alpha values could be used without impacting revenue. Search engines that experiment with higher alpha settings may be able to improve the relevance of their advertising for searchers and the value of the advertising to advertisers while also maximizing long-term revenue.

Our work demonstrates that it is important to consider the impact bad advertisements may have on other advertisements – the pollution effect – and suggests that advertisements are irrelevant should pay an additional penalty – a pollution tax – beyond what most search engines currently charge. We presented a simple but plausible model for the pollution effect, showed how changing the alpha to higher values can represent a pollution tax by changing irrelevant advertisements more, and showed how including the pollution effect can result in dramatically different conclusions about the optimal ranking of advertisements.

In the future, we would like to gather detailed empirical data on the pollution effect to better understand the long-term costs of showing less relevant advertising. We hope this data could be used for more sophisticated models of the pollution effect, especially for higher alpha settings, and allow us to better determine optimal rank ordering of advertising. In particular, we suspect that we may be able to narrow the range of alpha that optimizes revenue and may be able to develop a more detailed understanding of how to balancing the goals of the search engine (revenue), the searchers (relevance), and the advertisers (efficiency - revenue).

Finally, we would like to confirm the conclusions of our model on a live advertising engine. Our model makes several abstractions and assumptions – including that the CTR is an accurate measure of relevance for searchers, advertisers know their value and always bid rationally, and an accurate estimate of CTR is always known – that can be violated in practice. With live tests, we may be able to discover flaws in our current model that could be addressed in new models.

## 10. ACKNOWLEDGMENTS
Our thanks to Mukund Narasimhan from Microsoft Live Search for his help implementing the copulas used in the simulation, Sébastien Lahaie from Yahoo Research for providing additional details on the Monte Carlo simulations used in the previous work, Aleksander Kolcz from Microsoft Live Labs for helpful discussions, and to Microsoft for supporting this research.